\documentclass[aps,prd,reprint,onecolumn,a4paper,floatfix,nofootinbib,longbibliography]{revtex4-2}
\usepackage[a4paper,margin=18mm]{geometry}
\usepackage{float}
\usepackage{etoolbox}
\usepackage[T1]{fontenc}
\usepackage{lmodern,amsmath,amssymb,graphicx,booktabs,array,microtype}
\usepackage[colorlinks=true,linkcolor=blue,citecolor=blue,urlcolor=blue]{hyperref}
\numberwithin{equation}{section}

\newcommand{\BH}{\mathrm{BH}}
\newcommand{\ADM}{\mathrm{ADM}}
\newcommand{\HH}{\mathrm{H}}

\newcommand{\Sch}{\mathrm{Sch}}
\newcommand{\dd}{\mathrm{d}}
\newcommand{\ii}{i}
\newcommand{\rin}{r_{\mathrm{in}}}

\newcommand{\Mh}{M_{\mathrm{halo}}}

\begin{document}
\raggedbottom
\renewcommand{\topfraction}{.95}
\renewcommand{\bottomfraction}{.85}
\renewcommand{\textfraction}{.05}
\renewcommand{\floatpagefraction}{.8}
\setcounter{topnumber}{2}
\setcounter{totalnumber}{3}
\makeatletter
\def\@email#1#2{\endgroup\@AF@join{#1#2}}
\patchcmd{\frontmatter@abstractfont}{\small}{\normalsize}{}{\PackageError{SRDMBH}{Abstract font patch failed}{Check the REVTeX abstract font definition.}}
\patchcmd{\section}{0.8cm \@plus1ex \@minus .2ex}{14pt plus 2pt minus 2pt}{}{}
\patchcmd{\section}{0.5cm}{8pt}{}{}
\patchcmd{\subsection}{.8cm \@plus1ex \@minus .2ex}{11pt plus 2pt minus 2pt}{}{}
\patchcmd{\subsection}{.5cm}{6pt}{}{}
\makeatother
\setlength{\abovedisplayskip}{8pt plus 2pt minus 2pt}
\setlength{\belowdisplayskip}{8pt plus 2pt minus 2pt}
\setlength{\abovedisplayshortskip}{3pt plus 1pt}
\setlength{\belowdisplayshortskip}{5pt plus 2pt minus 2pt}
\title{Quasinormal Modes of Slowly Rotating Black Holes in Hernquist Halos: Redshift and Angular Momentum}
\author{Xing-Hui Feng}
\email{xhfeng@tju.edu.cn}
\author{Bo Wang}
\affiliation{Center for Joint Quantum Studies and Department of Physics, School of Science, Tianjin University, Tianjin 300350, China}

\begin{abstract}
We investigate axial quasinormal modes of black holes embedded in Hernquist halos, using an established analytic static geometry with an inner vacuum cavity extending to $4M_\BH$ and solving for frame dragging to first order in rotation. We compute the fundamental and first overtone branches for mass ratios $\epsilon\equiv M_\BH/\Mh$ of $0.1$ and $0.01$, and halo compactness $\mathcal C\equiv\Mh/a_0$ in the range $10^{-4}\leq\mathcal C\leq0.3$. The cavity geometry gives a local Kerr estimate in which redshift rescales the vacuum frequencies, horizon angular momentum sets the rotational splitting, and uniform frame dragging adds a real frequency shift. At zero spin, the cavity redshift reproduces the complex $\ell=2$ frequencies with relative residuals below $0.04\%$ for the fundamental and $0.31\%$ for the first overtone. Its leading weak compactness behavior, $1-\mathcal C$, explains the weak mass ratio dependence in dilute halos; at larger compactness, the exact redshift accounts for most of the separation between the two mass ratio sequences. At fixed total angular momentum, halos with vanishing local angular momentum and those with vanishing coordinate angular velocity exhibit opposite compactness trends in the real spin coefficient. At $\mathcal C=0.3$, their real fundamental mode spin coefficients differ by approximately $38\%$ and $54\%$ for the two mass ratios. Fixing the horizon angular momentum reduces these differences below $1\%$ for both overtones, showing that angular momentum partition accounts for most of the configuration dependence.
\end{abstract}
\maketitle
\section{Introduction}
Quasinormal modes (QNMs) characterize the damped response of perturbed black holes and provide a basis for inferring their properties from gravitational wave ringdown~\cite{BertiReview2009}. Matter surrounding a black hole can alter this spectrum by changing both the strong-field geometry and the propagation of perturbations through the exterior spacetime. Estimates for astrophysical environments find that their ringdown signatures can be difficult to distinguish from vacuum signals~\cite{Spieksma2025}. Early perturbative studies of black holes with external matter~\cite{Leung1999} and broader surveys of environmental effects~\cite{Barausse2014} established the importance of treating these contributions together. Self-consistent black-hole--halo models based on anisotropic matter have made it possible to investigate these effects within a common gravitational background~\cite{Cardoso2022,Figueiredo2023}. Analytic solutions with an inner vacuum region were obtained by Shen et al.~\cite{Shen2024Models}; their subsequent extension allows the inner edge of a Hernquist halo to be varied~\cite{Shen2024Inner}. We use this vacuum cavity geometry as the nonrotating limit of the backgrounds studied here.

The QNM spectrum depends on the halo background and on the dynamics of its matter, including the equation of state~\cite{Zhao2024}. For dilute halos, the leading axial QNM shift is a common redshift governed by the environmental potential~\cite{Pezzella2025}. The dependence on halo parameters has also been studied through shadows and QNM spectra in Schwarzschild--Hernquist backgrounds~\cite{FengZhang2025,QiCai2026}. Even on a given background, different prescriptions for matter perturbations can yield different spectra~\cite{Chakraborty2025}. The leading common redshift persists in models with nonzero radial pressure~\cite{Chen2026}, making it a natural reference against which to measure the remaining spectral shifts.

Rotating black-hole--halo models have been developed with different treatments of the surrounding matter. Fernandes and Cardoso constructed rotating black holes supported by anisotropic matter beyond the slow-rotation approximation~\cite{Fernandes2025}. Datta and Singha developed a slow-rotation treatment that incorporates the motion of the surrounding fluid and compared the resulting geodesic and epicyclic observables~\cite{Datta2026}. For the two fluid motions considered here, equal total angular momentum does not imply equal horizon angular momentum. Their axial QNM splitting therefore provides a way to distinguish angular momentum partition from other environmental contributions.

We compute the fundamental and first overtone axial gravitational QNMs to first order in rotation, using the perturbation framework of Refs.~\cite{PaniSlow2012,Pani2013,FengPeng2024} with a specified matter perturbation closure. At zero spin, we test the accuracy of the cavity redshift approximation for low multipoles and quantify the remaining mode-dependent shifts. For the rotational response, we compare halo configurations at fixed total and fixed horizon angular momentum. A local Kerr description of the vacuum cavity then separates the contributions from redshift, horizon spin, and uniform frame dragging; its discrepancy with the full QNM spectrum measures the residual effect of the exterior problem. Sections~\ref{sec:background} and \ref{sec:perturbations} present the halo backgrounds and the axial QNM calculation. Section~\ref{sec:predictions} derives the analytic predictions, which are compared with the numerical spectra in Sec.~\ref{sec:results}. Section~\ref{sec:conclusions} presents our conclusions.

\section{Hernquist halo background geometry}\label{sec:background}
We consider a static, spherically symmetric anisotropic fluid background and include rotation to first order in angular momentum. The static geometry is the Hernquist-type solution of Refs.~\cite{Shen2024Models,Shen2024Inner}, constructed within the vanishing-radial-pressure description of black holes surrounded by galactic matter~\cite{Cardoso2022,Figueiredo2023}.

\subsection{Matter model and field equations}
To first order in rotation, the metric takes the form
\begin{equation}
\dd s^2=-A\dd t^2+\frac{\dd r^2}{B}+r^2\dd\Omega_2^2
 -2r^2\sin^2\theta\,\chi\,\dd t\dd\phi,
\qquad B=1-\frac{2M(r)}{r}.
\end{equation}
The dragging function $\chi$ is first order in angular momentum; $A$, $B$, and the density retain their static forms to this order. Here $M(r)$ is the enclosed mass, and $\chi$ is the angular velocity of local zero-angular-momentum observers relative to infinity. The stress tensor is
\begin{equation}
T_{\mu\nu}=\rho u_\mu u_\nu+P_r k_\mu k_\nu+P_t\Pi_{\mu\nu},\qquad
\Pi_{\mu\nu}=g_{\mu\nu}+u_\mu u_\nu-k_\mu k_\nu,
\end{equation}
where $u^\mu=A^{-1/2}(1,0,0,\Omega)$ and $k^\mu=\sqrt B(0,1,0,0)$. The quantities $\rho$, $P_r$, and $P_t$ denote the rest-frame energy density, radial pressure, and tangential pressure, respectively, and $\Omega=\dd\phi/\dd t$ is the fluid angular velocity. We impose $P_r=0$, following the effective anisotropic fluid description motivated by an Einstein cluster~\cite{Cardoso2022}. Tangential stress then supports the prescribed matter distribution against gravity. The independent static Einstein equations and radial stress conservation yield
\begin{equation}
 M'=4\pi r^2\rho,\qquad \frac{A'}{A}=\frac{2M}{r(r-2M)},\qquad
 P_t=\frac{M\rho}{2(r-2M)}.\label{eq:static}
\end{equation}
Once $\rho(r)$ and the central mass are specified, these relations determine $M(r)$, $A(r)$, and $P_t(r)$. In particular, the tangential pressure follows from equilibrium rather than from an independently chosen equation of state.

Defining $\mathfrak j=\sqrt{B/A}$, the first-order equation for frame dragging can be written as~\cite{Datta2026}
\begin{equation}
 \frac{1}{r^4}(r^4\mathfrak j\chi')'
 +\frac{4}{r}\mathfrak j'\left(1+\frac{P_t}{\rho}\right)(\chi-\Omega)=0.
 \label{eq:rotation}
\end{equation}
With the static geometry fixed, Eq.~\eqref{eq:rotation} is a linear equation for $\chi$ once a relation between $\Omega$ and $\chi$ is prescribed.

\subsection{Hernquist profile and analytic static metric}
The Hernquist profile has an $r^{-1}$ inner cusp and an $r^{-4}$ outer tail, with a finite total mass~\cite{Hernquist1990}. Near a central black hole, however, the density need not follow the unmodified cusp. Black hole growth can produce a dark matter spike, while capture of particles on low-angular-momentum orbits depletes its innermost region. In relativistic collisionless spike calculations, this depletion can lead to a density cutoff at $r=4M_\BH$~\cite{Sadeghian2013}. The corresponding Schwarzschild orbit is marginally bound, whereas stable timelike circular orbits exist only for $r\geq6M_\BH$. Thus the motivation for an inner cutoff comes from particle capture, not from identifying $4M_\BH$ as a universal stability boundary.

To model such an inner depletion, we adopt the analytic Hernquist-type mass profile of Refs.~\cite{Shen2024Models,Shen2024Inner}, which allows the inner edge to be prescribed independently:
\begin{equation}
 M(r)=M_\BH+\Mh\frac{(r-\rin)^2}{(r+a_0)^2}\Theta(r-\rin).
 \label{eq:hqmass}
\end{equation}
Here $\Theta$ is the Heaviside step function, $\Mh$ is the total halo mass, $a_0$ is its scale radius, and $\rin>2M_\BH$ is the prescribed inner edge. As $r\to\infty$, $M(r)$ approaches $M_\ADM=M_\BH+\Mh$, the ADM mass. Outside the vacuum cavity, the density is
\begin{equation}
 \rho(r)=\frac{\Mh(a_0+\rin)(r-\rin)}{2\pi r^2(r+a_0)^3},\qquad r\geq\rin,
 \label{eq:hqdensity}
\end{equation}
and vanishes inside it. The density approaches zero continuously at $\rin$ and retains the Hernquist outer tail. For the QNM calculations, we take $\rin=4M_\BH$. The photon sphere at $r=3M_\BH$ then lies inside the vacuum cavity.

The analytic lapse~\cite{Shen2024Models,Shen2024Inner} can be expressed in terms of the roots $x_j$ of
\begin{equation}
 P(x)=x^3-(a_0+2M_\BH+2\Mh)x^2+4\Mh(a_0+\rin)x-2\Mh(a_0+\rin)^2.
\end{equation}
For simple roots, with $c_j=x_j^2/P'(x_j)$, it is
\begin{equation}
 A_{\rm ext}(r)=\exp\!\left[\sum_{j=1}^3c_j\log\left(1+\frac{a_0-x_j}{r}\right)\right].
 \label{eq:analyticA}
\end{equation}
The integration constant is chosen so that $A(r)\to1$ as $r\to\infty$. This identifies $t$ with the proper time of static observers at infinity and fixes the time normalization of the reported frequencies. The branches are continued from infinity, pairing complex conjugate roots to obtain a real, positive lapse wherever $r>2M(r)$; repeated roots are handled by continuity. This exterior condition is satisfied throughout the parameter scan, as discussed in Sec.~\ref{sec:light_rings}.

Matching the interior and exterior metrics at $\rin$ gives
\begin{equation}
 A_{\rm int}(r)=s^2\left(1-\frac{2M_\BH}{r}\right),\qquad
 s^2=\frac{A_{\rm ext}(\rin)}{1-2M_\BH/\rin}.
 \label{eq:staticcavity}
\end{equation}
For positive halo mass, $s<1$ and represents the redshift of the vacuum cavity relative to infinity. Because both $M$ and $M'$ are continuous at $\rin$, the metric and its first radial derivatives match without a surface layer. The horizon remains at $r=2M_\BH$ through first order in rotation.

\subsection{Frame dragging for the two fluid configurations}
We choose a nonrotating frame at infinity and fix the total angular momentum $J_\ADM$ by imposing
\begin{equation}
 \chi(r)=\frac{2J_\ADM}{r^3}+o(r^{-3}),\qquad r\to\infty.
 \label{eq:rotationboundary}
\end{equation}
Following Ref.~\cite{Datta2026}, we compare $\Omega=\chi$ and $\Omega=0$ on the same static background. Their physical distinction follows from
\begin{equation}
 u_\phi=\frac{r^2\sin^2\theta}{\sqrt A}(\Omega-\chi)
\end{equation}
to first order in rotation. When $\Omega=\chi$, the fluid has zero local angular momentum, $u_\phi=0$, and follows the dragging of local inertial frames. We refer to this as the ZAMO configuration. Equation~\eqref{eq:rotation} then integrates to
\begin{equation}
r^4\mathfrak j\chi'=-6J_\ADM,\qquad
\chi(r)=6J_\ADM\int_r^\infty\frac{\dd R}{R^4\mathfrak j(R)}.
\label{eq:zamo}
\end{equation}
In the Static configuration, $\Omega=0$. Introducing $\xi=r^4\mathfrak j\chi'$ and $\eta=r^3\mathfrak j\chi$, the dragging equation becomes
\begin{equation}
 \xi'=4\left(1+\frac{M}{2(r-2M)}\right)\frac{M'}{r-2M}\eta,\qquad
 \eta'=\frac{\xi}{r}+\left(\frac{3}{r}-\frac{M'}{r-2M}\right)\eta.
 \label{eq:staticspin}
\end{equation}
The asymptotic conditions are $(\xi,\eta)\to(-6J_\ADM,2J_\ADM)$ as $r\to\infty$. Here ``Static'' refers to the fluid's coordinate angular velocity, not to the spacetime: the fluid remains at fixed azimuth relative to infinity, but its covariant angular momentum is negative wherever $\chi>0$.

At first order, the Komar angular momentum enclosed by a spherical surface is
\begin{equation}
 J(r)=-\frac{r^4\mathfrak j\chi'}{6}=-\frac{\xi}{6}.
 \label{eq:komar}
\end{equation}
In the vacuum cavity, $J(r)$ is constant and equals the horizon angular momentum $J_\HH$. Consequently, $J_\HH=J_\ADM$ in the ZAMO configuration. In the Static configuration, the matter contribution is negative for positive dragging and $\rho+P_t>0$, giving $J_\HH>J_\ADM$ for the models considered here.

We determine $\chi$ from the analytic static background: by quadrature of Eq.~\eqref{eq:zamo} for ZAMO matter and by inward integration of Eq.~\eqref{eq:staticspin} for Static matter. At the cavity edge, $\chi$ and $r^4\mathfrak j\chi'$ are continuous, excluding a surface layer of angular momentum. Because the equation is linear, the result can be normalized to any sufficiently small $J_\ADM$. The two configurations therefore have the same density and static redshift but different distributions of angular momentum between the horizon and the halo.

\section{Axial perturbations and numerical formulation}\label{sec:perturbations}
We calculate axial QNMs for both backgrounds, imposing ingoing waves at the horizon and outgoing waves at infinity.

\subsection{Axial perturbations and spectral formulation}
We express lengths and times in units of $M_\BH$ and frequencies and angular velocities in units of $M_\BH^{-1}$. Thus $\hat r=r/M_\BH$, $\hat t=t/M_\BH$, $\hat\omega=M_\BH\omega$, and $\hat\chi=M_\BH\chi$. We subsequently omit hats and set $M_\BH=1$, so our choice $\rin=4M_\BH$ corresponds to $r=4$.

For the decoupled axial gravitational branch of the anisotropic fluid system~\cite{FengPeng2024}, the master equation is
\begin{equation}
 \frac{\dd^2\Psi}{\dd r_*^2}+[\omega^2-V_0-mV_\chi]\Psi=0,
 \qquad \frac{\dd r_*}{\dd r}=\frac{1}{\sqrt{AB}},
 \label{eq:master}
\end{equation}
where
\begin{align}
 V_0&=\frac{A}{r^2}\left[\ell(\ell+1)-\frac{6M}{r}+M'\right],\nonumber\\
 V_\chi&=2\omega\chi+
 \frac{2A}{\ell(\ell+1)\omega r^2}
 \left\{r(r-2M)\chi''-[r(2+M')-6M]\chi'\right\}.
 \label{eq:rotpotential}
\end{align}
Following Ref.~\cite{FengPeng2024}, we assume that the fluid perturbations produce no radial--tangential shear stress in the local rest frame and eliminate the constrained fluid response to obtain the axial master equation. At first order, the frequency shift can be determined within a fixed-$\ell$ axial sector, although the complete rotating perturbation system couples different sectors~\cite{PaniSlow2012,Pani2013}.

For the time dependence $e^{-\ii\omega t}$, the QNM boundary conditions are
\begin{equation}
 \Psi\sim e^{-\ii(\omega-m\chi_\HH)r_*}\quad(r\to2),\qquad
 \Psi\sim e^{+\ii\omega r_*}\quad(r\to\infty),
\end{equation}
where $\chi_\HH$ is the horizon angular velocity at this order. At the cavity edge, the master equation requires continuity of $\Psi$ and $\dd\Psi/\dd r_*$. Since $\sqrt{AB}$ is continuous, these conditions are equivalent to matching $\Psi$ and $\Psi'$ at $r=4$.

We define the total and horizon spin parameters by
\begin{equation}
 j_\ADM=\frac{J_\ADM}{M_\BH^2},\qquad
 \kappa=\frac{J_\HH}{J_\ADM},\qquad j_\HH=\kappa j_\ADM.
 \label{eq:spinlabels}
\end{equation}
Both spin parameters use the central black hole mass as their normalization. The ratio $\kappa$ describes angular momentum partition and may exceed unity if the halo contributes negative angular momentum. We expand the frequencies as
\begin{equation}
 \omega_{\ell mn}=\omega_{0,\ell n}+m j_\ADM\omega_{1,\ell n}
 +O(j_\ADM^2).
 \label{eq:expansion}
\end{equation}
The zero-spin coefficient $\omega_0$ is the same for both configurations; $\omega_1$ gives their respective splittings at fixed total spin. At fixed horizon spin, we denote the corresponding slope by $\omega_{1,X}^{\mathrm{H}}\equiv\omega_{1,X}^{\rm ADM}/\kappa_X$, where $X=Z,S$ labels the ZAMO and Static configurations. At this order the coefficient applies to all allowed $m$, with no linear correction for $m=0$. The expansion assumes that both $|j_\ADM|$ and $|j_\HH|$ remain small.

\subsection{Numerical method}
We solve the radial boundary-value problem by matching the logarithmic derivative $F=\Psi'/\Psi$ at the cavity edge. The interior solution is initialized by an ingoing Frobenius series at the horizon; the exterior solution is integrated inward from an outgoing boundary condition. With $g=\sqrt{AB}$, the master equation becomes
\begin{equation}
 F'=-F^2-\frac{g'}{g}F-\frac{\omega^2-V_0-mV_\chi}{AB}.
\end{equation}
The QNM condition is $\Delta F\equiv F_{\rm in}-F_{\rm out}=0$ at $r=4$. After solving $\Delta F(\omega,0)=0$ for $\omega_0$, we differentiate the matching condition with respect to $z=mj_\ADM$ to obtain
\begin{equation}
 \omega_1=-\left.\frac{\partial_z\Delta F}{\partial_\omega\Delta F}\right|_{\omega=\omega_0,\,z=0}.
 \label{eq:adjoint}
\end{equation}
These derivatives are computed from the differentiated radial equation and boundary data, including the spin dependence of the ingoing exponent. We evaluate the static geometry analytically and obtain dragging derivatives from the rotational field equation. Each mode is continuously tracked from its Schwarzschild counterpart.

For the outer integration we use the complex contour $r=4+e^{\ii\vartheta}(e^v-1)$, $v\geq0$, choosing $\operatorname{Im}(\omega e^{\ii\vartheta})>0$ so that the outgoing wave decays along the contour. At a finite outer endpoint, the primary calculation uses $F=\ii\omega(1+2M_\ADM/r)$. As an independent boundary test, we impose the local WKB condition
\begin{equation}
 F_{\rm WKB}=\frac{\ii\omega}{g}\sqrt{1-\frac{V}{\omega^2}}
              +\frac{V'}{4(\omega^2-V)},\qquad V=V_0+mV_\chi,
 \label{eq:wkbBC}
\end{equation}
The square-root branch is continued from the outgoing vacuum solution. This prescription incorporates the endpoint metric and potential, including the first WKB amplitude correction, and tests the asymptotic initializer even when the endpoint lies within the halo's scale radius.

We cross-check the frequencies using a two-domain finite difference matrix method. Radial derivatives are approximated by local Taylor finite difference stencils~\cite{Fornberg1988}; the nodal values then satisfy a matrix eigenvalue problem, building on the interpolation and QNM matrix formulations of Refs.~\cite{LinQian2016,LinQian2017}. The domains meet at the halo inner edge, where the wave function and its radial derivative are matched. Grid refinement and changes of stencil width test convergence across the full dataset. Selected modes are also checked using Chebyshev spectral collocation~\cite{Trefethen2000,Jansen2017}. These independent radial discretizations share the background, potential, and original asymptotic boundary condition with the matching calculation; the WKB test probes the boundary approximation itself.

\section{Vacuum cavity and analytic predictions}\label{sec:predictions}
Inside the vacuum cavity, the geometry is Kerr-like at first order in spin, apart from the time normalization and a constant dragging offset. The corresponding change of coordinates relates the local mode frequencies to those measured at infinity.

\subsection{Local Kerr geometry and frequency estimates}\label{sec:cavity}
In $2M_\BH<r<\rin$, set $M_\BH=1$ and write
\begin{equation}
 B=f=1-\frac2r,\qquad A=s^2f,\qquad
 \chi=J_\ADM\left(c+\frac{2s\kappa}{r^3}\right),\qquad s>0.
 \label{eq:cavity}
\end{equation}
The exterior solution fixes $s$, $c$, and $\kappa$. The coordinate transformation $T=st$, $\Phi=\phi-cJ_\ADM t$ brings the cavity metric to the Kerr form at linear order in the horizon angular momentum $J_\HH$. A perturbation proportional to $e^{-\ii\omega t+\ii m\phi}$ therefore has frequency $(\omega-mcJ_\ADM)/s$ with respect to $T$. This observation motivates the local estimates
\begin{equation}
 \omega_0^{\mathrm{loc}}=s\omega_\Sch,\qquad
 \omega_1^{\mathrm{loc}}=s\kappa\omega_{1,\mathrm{Kerr}}+c.
 \label{eq:localestimate}
\end{equation}
The factor $s$ rescales both the oscillation frequency and its damping rate, whereas $\kappa$ accounts for the horizon spin at fixed total angular momentum. The constant $c$ produces an additive real shift $mJ_\ADM c$: although removable by a local redefinition of azimuth, it represents frame dragging relative to the reference frame that is nonrotating at infinity. Equation~\eqref{eq:localestimate} retains the finite-$\ell$ vacuum QNM frequencies, including the complex Kerr spin coefficient. It remains an approximation to the global QNM spectrum because the outgoing boundary condition is imposed beyond the cavity, where the effective potential differs from Kerr. We quantify the departure from the local rotational estimate by defining
\begin{equation}
 \delta\omega_{1,X}=\omega_{1,X}^{\rm ADM}
 -\left(s\kappa_X\omega_{1,\mathrm{Kerr}}+c_X\right),
 \qquad X=Z,S.
 \label{eq:spinresidual}
\end{equation}
This complex residual measures the correction from wave propagation through the halo beyond the redshift, horizon spin rescaling, and uniform frame dragging encoded in the local cavity geometry.

\subsection{Weak compactness predictions}\label{sec:weak}
We parameterize the halo by
\begin{equation}
 \mathcal C\equiv\frac{\Mh}{a_0},\qquad \epsilon\equiv\frac{M_\BH}{\Mh}.
 \label{eq:parameters}
\end{equation}
The compactness $\mathcal C$ characterizes the halo's potential depth, while $\epsilon$ is the black-hole-to-halo mass ratio. The dilute hierarchy $M_\BH\ll\Mh\ll a_0$ requires both parameters to be small. With $\rin=4M_\BH$, the cavity-to-halo scale ratio is $\rin/a_0=4\epsilon\mathcal C\ll1$ in this regime. At fixed $M_\BH$ and $\epsilon$, varying $\mathcal C$ changes $a_0$ while leaving both masses unchanged. At fixed $M_\BH$ and $\mathcal C$, varying $\epsilon$ changes the halo mass and scale radius together while preserving the characteristic potential scale.

The cavity redshift follows directly from the static field equations and the matching condition. Subtracting the Schwarzschild logarithmic derivative of the lapse and integrating with the normalization at infinity gives, in units $M_\BH=1$,
\begin{equation*}
 \ln s=-\int_4^\infty
 \frac{\Delta M(r)\,\dd r}{(r-2)[r-2-2\Delta M(r)]},
 \qquad \Delta M(r)=\Mh\frac{(r-4)^2}{(r+a_0)^2}.
\end{equation*}
At fixed finite $\epsilon>0$, its expansion as $\mathcal C\to0$ yields
\begin{equation}
 s=1-\mathcal C+\mathcal C^2
 \left[\frac16-2\epsilon+4\epsilon\ln\frac1{2\epsilon\mathcal C}\right]
 +o(\mathcal C^2).
 \label{eq:redshiftsecond}
\end{equation}
The leading shift depends only on compactness. The next terms contain a logarithm, so at finite mass ratio the expansion is not simply polynomial in $\mathcal C$. At the level of the displayed $\mathcal C^2$ coefficient, the small mass ratio limit agrees with Ref.~\cite{FengZhang2025}: at fixed $\mathcal C>0$, $\epsilon\ln[1/(2\epsilon\mathcal C)]\to0$ as $\epsilon\to0$, leaving $1/6$. The finite-$\epsilon$ terms depend on the background. In Ref.~\cite{FengZhang2025} the halo reaches the horizon, whereas our model contains a vacuum cavity. The logarithmic term in Eq.~\eqref{eq:redshiftsecond} is part of the redshift and is removed from the spectral residual when the exact $s$ is divided out.

At fixed finite $\epsilon$, the same weak compactness limit sends $a_0\to\infty$. In the region controlling the leading halo integral, $\eta/J_\ADM=2+o(1)$, and the Static dragging equation gives $\xi'/J_\ADM\simeq8M'/r$. Since $\int_{\rin}^\infty M'(r)\,\dd r/r=\mathcal C+o(\mathcal C)$, inward integration from $\xi(\infty)/J_\ADM=-6$ yields
\begin{equation}
 \kappa_{\rm Static}=1+\frac43\mathcal C+o(\mathcal C),\qquad
 \kappa_{\rm ZAMO}=1,\qquad c=o(\mathcal C).
\end{equation}
Equation~\eqref{eq:localestimate} then predicts
\begin{equation}
 \frac{\omega_0}{\omega_{\rm Sch}}\simeq1-\mathcal C,\qquad
 \frac{\omega_{1,\rm ZAMO}^{\rm ADM}}{\omega_{1,\rm Kerr}}\simeq1-\mathcal C,\qquad
 \frac{\omega_{1,\rm Static}^{\rm ADM}}{\omega_{1,\rm Kerr}}\simeq1+\frac13\mathcal C.
 \label{eq:weakspin}
\end{equation}
The opposite trends at fixed total spin arise from competition between redshift and angular momentum partition. For ZAMO matter, $\kappa=1$, and redshift suppresses the local spin response. In the Static configuration, the increase in $J_\HH$ more than compensates for this suppression, giving $s\kappa=1+\mathcal C/3+o(\mathcal C)$. When horizon spin is fixed, dividing by $\kappa$ removes the leading partition effect, leaving a factor $1-\mathcal C$ in both rotational coefficients; the uniform dragging term is subleading.

\subsection{Light rings and the eikonal limit}\label{sec:light_rings}
The same cavity geometry also determines the large-$\ell$ light ring prediction. We first establish that the relevant null circular orbit lies inside the vacuum region. At zero spin, the static field equations imply the circular orbit condition $r_c=3M(r_c)$. For $r\geq4$, the following sufficient bound excludes additional exterior light rings:
\begin{equation}
 \frac{M(r)}r=\frac1r+\Mh\frac{(r-4)^2}{r(r+a_0)^2}
 \leq\frac14+\Mh\frac{r}{(r+a_0)^2}
 \leq\frac{1+\mathcal C}{4}<\frac13
 \quad(\mathcal C<1/3).
\end{equation}
For $\mathcal C<1/3$, the exterior has neither a circular null orbit satisfying $r=3M(r)$ nor a zero of $B$. Our scan ends at $\mathcal C=0.3$, so the only static light ring lies inside the cavity. The rotating orbit below is its continuous deformation at first order in spin.

For equatorial null geodesics, let $E=-p_t$ and $L=p_\phi$ be the conserved energy and angular momentum, and let a dot denote differentiation with respect to an affine parameter. The null condition gives, to first order in spin,
\begin{equation}
 \dot r^2=V_r=B\left(\frac{E^2}{A}-\frac{L^2}{r^2}-\frac{2EL\chi}{A}\right).
\end{equation}
Imposing $V_r(r_c)=V'_r(r_c)=0$ at fixed $E$ and $L$ and eliminating the impact parameter $L/E$ gives the circular orbit condition and corresponding angular frequency $\Omega_c=\dot\phi_c/\dot t_c$:
\begin{equation}
 r_c A'_c-2A_c+2\sigma r_c^2\sqrt{A_c}\chi'_c=0,\qquad
 \Omega_c^\sigma=\chi_c+\sigma\frac{\sqrt{A_c}}{r_c}.
\end{equation}
Here $\sigma=\pm1$ labels the two orbital directions, and a subscript $c$ denotes evaluation at $r_c$. Linearizing the radial motion gives the instability exponent measured by an observer at infinity, $\lambda_c^2=V''_r(r_c)/(2\dot t_c^2)$~\cite{Cardoso2009}. Using the circular orbit condition, this becomes
\begin{equation}
 (\lambda_c^\sigma)^2=B_c\left[\frac{A_c}{r_c^2}-\frac{A''_c}{2}
 -\sigma\sqrt{A_c}(4\chi'_c+r_c\chi''_c)\right].
 \label{eq:lambda}
\end{equation}
At zero spin, the expression reduces to
\begin{equation}
 \lambda_c^2=\frac{A_c}{r_c^2}[1-3M'(r_c)].
\end{equation}

For the present cavity, the zero-spin light ring is at $r_c=3$. Substituting Eq.~\eqref{eq:cavity} gives, through first order in rotation,
\begin{align}
 r_c^\sigma&=3-\sigma\frac{2\kappa J_\ADM}{\sqrt3}+O(J_\ADM^2),\\
 \Omega_c^\sigma&=\sigma\frac{s}{3\sqrt3}
 +J_\ADM\left(c+\frac{2s\kappa}{27}\right)+O(J_\ADM^2),\\
 \lambda_c^\sigma&=\frac{s}{3\sqrt3}+O(J_\ADM^2).
 \label{eq:lrinterior}
\end{align}
Rotation shifts the orbital frequency but leaves the instability exponent unchanged at first order, apart from its static redshift factor. Indeed, $4\chi'+r\chi''=0$ inside the cavity, and $f/r^2$ is stationary at $r=3$. The explicit dragging term and the first-order orbit shift therefore make no linear contribution to $\lambda_c$. This cancellation follows from the vacuum interior and need not hold for a light ring embedded in matter.

The standard geodesic correspondence~\cite{Cardoso2009} consequently gives, at fixed overtone and large $\ell$,
\begin{equation}
 \omega_{\ell mn}\simeq
 \left(\ell+\tfrac12\right)\frac{s}{3\sqrt3}
 +mJ_\ADM\left(c+\frac{2s\kappa}{27}\right)
 -\ii\left(n+\tfrac12\right)\frac{s}{3\sqrt3}.
 \label{eq:eikonal}
\end{equation}
This is also the eikonal limit of Eq.~\eqref{eq:localestimate}, since the vacuum coefficients satisfy
\begin{equation*}
 \omega_{\Sch,\ell n}\simeq
 \frac{(\ell+\tfrac12)-\ii(n+\tfrac12)}{3\sqrt3},
 \qquad \omega_{1,\mathrm{Kerr},\ell n}\longrightarrow\frac{2}{27}.
\end{equation*}
The geodesic approximation captures the leading eikonal terms, while Eq.~\eqref{eq:localestimate} retains the finite multipole vacuum coefficients. In particular, the absence of a leading eikonal spin correction to damping does not imply $\operatorname{Im}\omega_1=0$ at finite $\ell$. Because $c$ is real, Eq.~\eqref{eq:spinresidual} gives
\begin{equation*}
 \operatorname{Im}\omega_1
 =s\kappa\operatorname{Im}\omega_{1,\mathrm{Kerr}}
 +\operatorname{Im}\delta\omega_1.
\end{equation*}
The first term is the redshifted finite-$\ell$ Kerr response; the second measures the departure of the global mode from the local approximation.

\section{Numerical results}\label{sec:results}
The spectra allow us to separate the common redshift from the rotational response and to measure the corrections left by the cavity approximation. We first specify the parameter range and vacuum benchmarks, then examine the common redshift, rotational splittings, and mode-dependent residuals.

\subsection{Common redshift and mass ratio dependence}
The two principal sequences use $\epsilon=0.1$ and $0.01$, corresponding to halo masses $10M_\BH$ and $100M_\BH$, respectively. For each mass ratio, we sample
\begin{equation}
 \mathcal C\in\{10^{-4},10^{-3},0.01,0.03,0.05,0.1,0.15,0.2,0.25,0.3\}.
\end{equation}
The resulting scale radii span $a_0/M_\BH=100/3$ to $10^6$. The maximum ratio $\rin/a_0$ is $0.12$ for $\epsilon=0.1$ and $0.012$ for $\epsilon=0.01$. The halo scale therefore remains larger than the cavity radius across the scan, which extends from dilute to moderately compact configurations. The larger compactnesses are included to probe the model rather than to represent typical galactic halos. We compute the $\ell=2$ fundamental and first overtone branches at all 20 $(\epsilon,\mathcal C)$ pairs, for both rotating fluid configurations. The Schwarzschild and slowly rotating Kerr benchmarks~\cite{HatsudaKimura2020} yield
\begin{align}
 \omega_{{\rm Sch},0}&=0.373672-0.088962\ii,&
 \omega_{1,{\rm Kerr},0}&=0.062883+0.000998\ii,\nonumber\\
 \omega_{{\rm Sch},1}&=0.346711-0.273915\ii,&
 \omega_{1,{\rm Kerr},1}&=0.071940+0.006385\ii.
\end{align}
The second subscript on the Kerr coefficient denotes the overtone number. Frequencies and spin coefficients are displayed to six decimal places, but all plots, ratios, and residuals use unrounded values.

\begin{figure}[!htb]\centering\includegraphics[width=.70\textwidth]{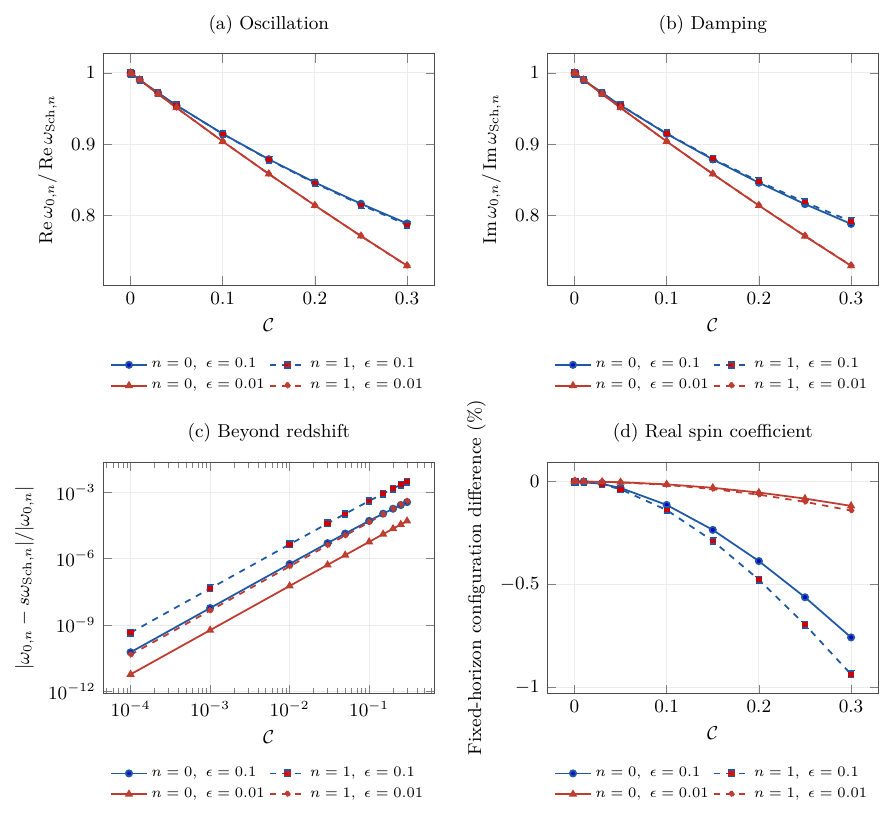}
\caption{The two $\ell=2$ overtones across the compactness scan. Panels (a) and (b) normalize each mode by its Schwarzschild value; the nearly coincident curves are distinguished by the residual in panel (c). Panel (d) shows the relative difference between the real spin coefficients of the two matter configurations at fixed horizon spin, $100[\operatorname{Re}\omega_{1,S}^{\mathrm{H}}/\operatorname{Re}\omega_{1,Z}^{\mathrm{H}}-1]$. Colors identify the mass ratio and line styles distinguish overtones.}\label{fig:combinedfreq}\end{figure}

As $\mathcal C$ increases, both the oscillation frequency and the magnitude of the damping rate decrease (Fig.~\ref{fig:combinedfreq}). Even for the two $\ell=2$ modes, the common redshift estimate $\omega_{0,n}\simeq s\omega_{{\rm Sch},n}$ accurately describes both parts of the complex frequency. Equation~\eqref{eq:redshiftsecond} explains the weak mass ratio dependence at low compactness: the leading redshift depends only on $\mathcal C$, while the $\epsilon$-dependent terms are subleading. In the numerical spectra, changing $\epsilon$ from $0.1$ to $0.01$ at $\mathcal C=0.01$ changes the real frequency by only about $0.02\%$ for either overtone.

At $\mathcal C=0.3$, decreasing $\epsilon$ from $0.1$ to $0.01$ reduces the real frequencies by $7.511\%$ and $7.316\%$ for $n=0$ and $1$, respectively, and the magnitudes of the damping rates by $7.458\%$ and $7.825\%$. Over the same change, the redshift falls from $s=0.788763$ to $0.729337$, a reduction of $7.534\%$. Hence redshift dominance does not imply mass ratio independence: the exact redshift itself carries substantial $\epsilon$ dependence at larger compactness. Table~\ref{tab:joint} lists representative complex frequencies and spin coefficients.
\begin{table}[H]\centering\small\renewcommand{\arraystretch}{1.18}\setlength{\tabcolsep}{5pt}
\caption{Representative $\ell=2$ frequencies and spin coefficients for $M_\BH=1$ and $\rin=4$. The spin coefficients multiply $mj_\ADM$.}\label{tab:joint}
\begin{tabular}{rrrccc}\toprule
$\epsilon$ & $\mathcal C$ & $n$ & $\omega_{0,n}$ & $\omega_{1,Z}^{\rm ADM}$ & $\omega_{1,S}^{\rm ADM}$ \\ \midrule
0.1 & 0.1 & 0 & $0.341732-0.081351\,\ii$ & $0.057563+0.000895\,\ii$ & $0.064962+0.001013\,\ii$ \\
0.1 & 0.1 & 1 & $0.316976-0.250632\,\ii$ & $0.066022+0.005829\,\ii$ & $0.074489+0.006591\,\ii$ \\
0.1 & 0.3 & 0 & $0.294651-0.070104\,\ii$ & $0.049909+0.000679\,\ii$ & $0.069088+0.000960\,\ii$ \\
0.1 & 0.3 & 1 & $0.272739-0.216838\,\ii$ & $0.058226+0.004955\,\ii$ & $0.080456+0.006958\,\ii$ \\
0.01 & 0.1 & 0 & $0.337731-0.080405\,\ii$ & $0.056841+0.000900\,\ii$ & $0.065126+0.001032\,\ii$ \\
0.01 & 0.1 & 1 & $0.313354-0.247584\,\ii$ & $0.065046+0.005770\,\ii$ & $0.074525+0.006612\,\ii$ \\
0.01 & 0.3 & 0 & $0.272521-0.064876\,\ii$ & $0.045907+0.000714\,\ii$ & $0.070695+0.001103\,\ii$ \\
0.01 & 0.3 & 1 & $0.252784-0.199870\,\ii$ & $0.052645+0.004653\,\ii$ & $0.081053+0.007178\,\ii$ \\
\bottomrule\end{tabular}
\end{table}

\subsection{Total versus horizon angular momentum}\label{sec:angular_partition}
\begin{figure}[!htb]\centering\includegraphics[width=.70\textwidth]{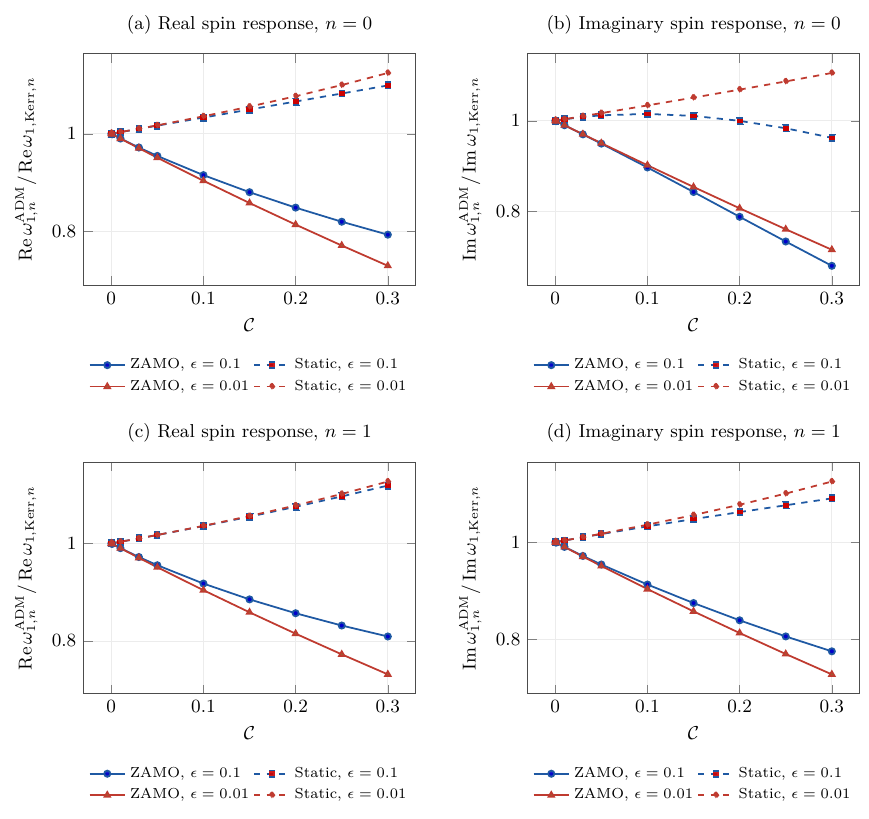}
\caption{Spin responses at fixed total angular momentum, normalized by the corresponding vacuum Kerr coefficient for each overtone. For the Static configuration with $\epsilon=0.1$, the imaginary coefficient of the fundamental turns over across the sampled points, whereas the first overtone coefficient continues to increase.}\label{fig:combinedspin}\end{figure}

Figure~\ref{fig:combinedspin} shows opposite compactness trends in the real spin coefficient at fixed $j_\ADM$: it decreases for ZAMO matter and increases for Static matter, for both overtones. The weak compactness factors $1-\mathcal C$ and $1+\mathcal C/3$ derived in Sec.~\ref{sec:weak} explain these opposite initial trends.

At $\mathcal C=0.3$, the Static configuration has $\kappa=1.394872$ for $\epsilon=0.1$ and $\kappa=1.541782$ for $\epsilon=0.01$. At fixed total spin, the differences in the real coefficients between the configurations are $38.429\%$ and $53.996\%$ for the fundamental, and approximately $38.18\%$ and $53.96\%$ for the first overtone. Dividing the coefficients by $\kappa$ reduces the magnitudes of these differences below $1\%$ in all four cases, as shown in Fig.~\ref{fig:combinedfreq}(d): the fundamental differences become $-0.7584\%$ and $-0.1184\%$, while those of the first overtone become approximately $-0.938\%$ and $-0.141\%$. The large separation at fixed total spin therefore primarily reflects a difference in horizon angular momentum. The smaller residual difference at fixed horizon spin contains the effects of exterior frame dragging and wave propagation. At fixed horizon spin, the local estimate retains the uniform dragging offset through $\omega_1^{\mathrm{loc}}/\kappa=s\omega_{1,\mathrm{Kerr}}+c/\kappa$.

\subsection{Residual scaling beyond redshift}\label{sec:residual_scaling}
We measure the departure of the zero-spin spectrum from common redshift scaling by defining the complex relative residual and its rescaled counterpart:
\begin{equation}
 \delta_n=\frac{\omega_{0,n}}{s\omega_{{\rm Sch},n}}-1,
 \qquad Q_n=\frac{\delta_n}{\epsilon\mathcal C^2}.
 \label{eq:scaledresidual}
\end{equation}
This complex normalization differs from the nonnegative residual $|\omega_{0,n}-s\omega_{{\rm Sch},n}|/|\omega_{0,n}|$ in Fig.~\ref{fig:combinedfreq}(c). For fixed $r>4$ with $r\ll a_0$, the density expands as
\begin{equation}
 \rho(r)=\frac{\epsilon\mathcal C^2(r-4)}{2\pi r^2}
 \left[1+O\bigl(\epsilon\mathcal C(r+4)\bigr)\right].
\end{equation}
The local source amplitude $\epsilon\mathcal C^2$ suggests the scaling to test for the global QNM residual, which includes contributions from both the exterior potential and boundary matching.

\begin{figure}[!htb]\centering\includegraphics[width=.70\textwidth]{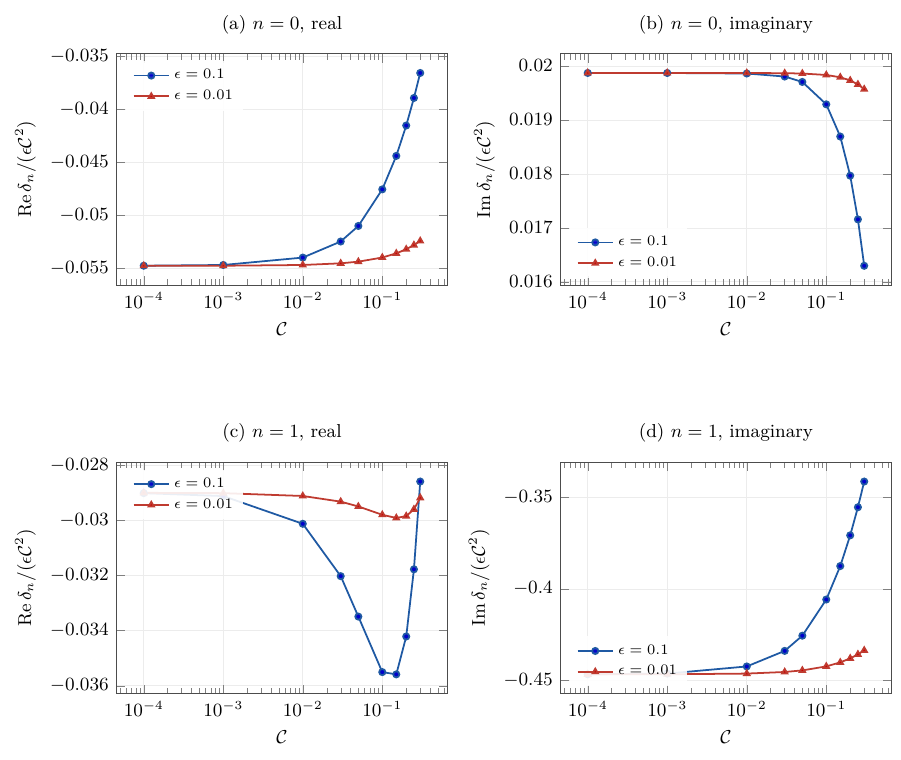}
\caption{The rescaled complex zero-spin residual $Q_n$ defined in Eq.~\eqref{eq:scaledresidual}. The imaginary part of the first overtone coefficient is substantially larger in magnitude. Colors distinguish mass ratios and the panels separate the real and imaginary parts.}\label{fig:scaling}\end{figure}

The estimated weak compactness limits are
\begin{equation}
 Q_0\simeq-0.0548+0.0199\ii,\qquad
 Q_1\simeq-0.0290-0.447\ii.
\end{equation}
We estimate the small-$\mathcal C$ coefficients by linear extrapolation of the $\epsilon=0.1$ results at $\mathcal C=10^{-4}$ and $10^{-3}$, and use the corresponding $\epsilon=0.01$ points as an independent check. The two mass ratios give nearly common limiting values (Fig.~\ref{fig:scaling}); the data do not resolve higher-order terms, including possible logarithms.

Across the full scan, the relative deviations in Fig.~\ref{fig:combinedfreq}(c) stay below $0.04\%$ for the fundamental and $0.31\%$ for the first overtone. The maximum first overtone residual is roughly eight times larger, yet remains small compared with the overall frequency shift. The same mode dependence is apparent in the redshift-independent two-mode ratio
\begin{equation}
 \mathcal R_{10}=\frac{\omega_{0,1}\omega_{{\rm Sch},0}}{\omega_{0,0}\omega_{{\rm Sch},1}}-1,
\end{equation}
whose magnitude at $\mathcal C=0.3$ is $3.22\times10^{-3}$ and $4.08\times10^{-4}$ for the two mass ratios.

\subsection{Decomposition of the Static damping response turnover}
\begin{figure}[!htb]\centering\includegraphics[width=.70\textwidth]{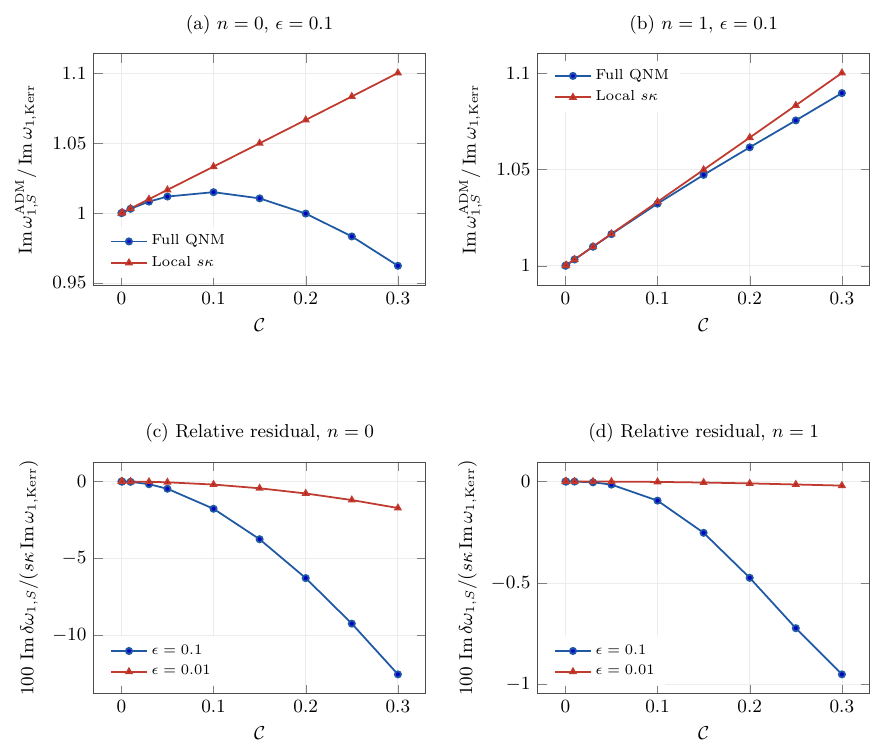}
\caption{Decomposition of the imaginary spin coefficient in the Static configuration. The upper panels compare $\operatorname{Im}\omega_{1,S}^{\rm ADM}/\operatorname{Im}\omega_{1,\mathrm{Kerr}}$ with the local prediction $s\kappa_S$ for $\epsilon=0.1$. The lower panels show the signed residual relative to the local term for both mass ratios. The fundamental mode turnover results from competition between the increasing local contribution and a negative exterior residual.}\label{fig:turnover}\end{figure}

A small zero-spin residual does not imply a small relative correction to the rotational damping response. The imaginary spin coefficient is itself small, so a modest departure from its local estimate can change its compactness trend. Since $c_S$ is real, Eq.~\eqref{eq:spinresidual} gives
\begin{equation}
 \operatorname{Im}\omega_{1,S}^{\rm ADM}
 =s\kappa_S\operatorname{Im}\omega_{1,{\rm Kerr}}
  +\operatorname{Im}\delta\omega_{1,S}.
 \label{eq:turnover}
\end{equation}
For $\epsilon=0.1$, the local contribution increases with $\mathcal C$ throughout the scan. In the fundamental branch, however, the residual is negative and grows sufficiently in magnitude to reverse the trend of the full coefficient (Fig.~\ref{fig:turnover}). From $\mathcal C=0.1$ to $0.3$, the local contribution rises from $0.001031$ to $0.001098$, while the residual changes from $-0.000018$ to $-0.000138$. Their sum consequently decreases from $0.001013$ to $0.000960$. The largest value on our grid occurs at $\mathcal C=0.1$.

For the first overtone at the same mass ratio, the corresponding relative residual is only $-0.0941\%$ at $\mathcal C=0.1$ and $-0.9503\%$ at $\mathcal C=0.3$, compared with $-1.7693\%$ and $-12.5306\%$ for the fundamental. It therefore does not overcome the growing local contribution, and the sampled first overtone imaginary coefficient continues to increase. For $\epsilon=0.01$, both sequences also increase over the sampled range. This turnover concerns the linear damping correction $mj_\ADM\operatorname{Im}\omega_1$; the zero-spin modes remain damped throughout the scan.

\subsection{Higher multipoles and the eikonal prediction}\label{sec:multipoles}
\begin{figure}[H]\centering\includegraphics[width=.70\textwidth]{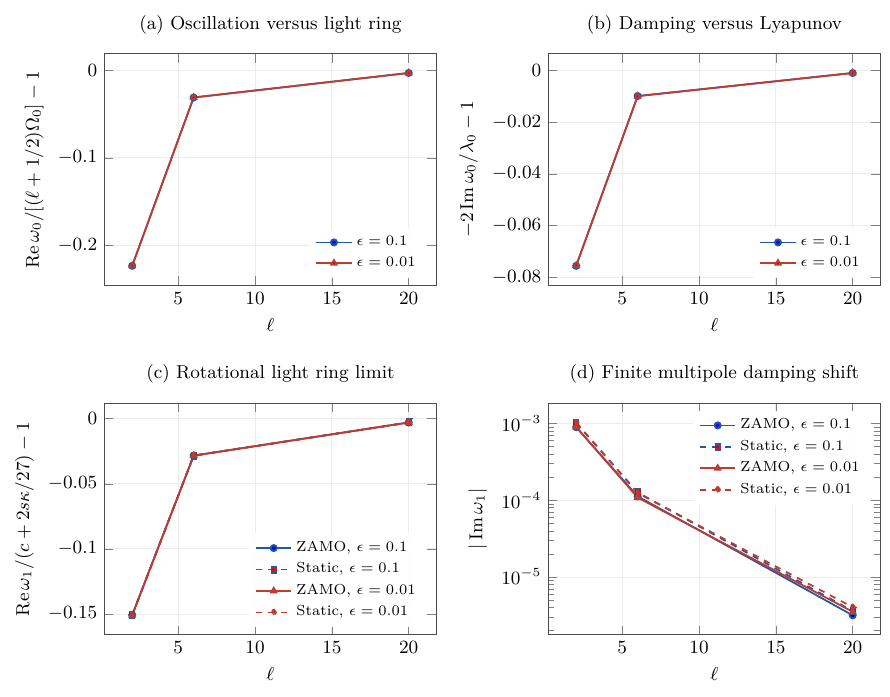}
\caption{Fundamental modes at $\mathcal C=0.1$. Panels (a)--(c) show the relative deviations from the light ring predictions, with $\Omega_0=\lambda_0=s/(3\sqrt3)$. Panel (d) displays the decrease of the imaginary spin coefficient with multipole order. Lines connect the computed values $\ell=2,6,20$.}\label{fig:multipolesC}\end{figure}

To compare with the geodesic limit, we also calculate the fundamental mode at $\ell=6$ and $20$ for both mass ratios at $\mathcal C=0.1$. The results appear in Table~\ref{tab:modesC}; the corresponding $\ell=2$ values are listed in Table~\ref{tab:joint}. Figure~\ref{fig:multipolesC} compares $\operatorname{Re}\omega_0/(\ell+1/2)$, $-\operatorname{Im}\omega_0/(n+1/2)$, and $\omega_1$ with the geodesic predictions derived in Sec.~\ref{sec:light_rings}. Because the relevant light ring lies inside the vacuum cavity, the real rotational coefficient approaches $c+2s\kappa/27$ at high multipole order, while its imaginary part decreases toward zero over the sampled multipoles.
\begin{table}[H]\centering\small\renewcommand{\arraystretch}{1.18}\setlength{\tabcolsep}{5pt}\caption{Higher multipole fundamental modes at $\mathcal C=0.1$, continuously tracked from their Schwarzschild counterparts.}\label{tab:modesC}
\begin{tabular}{rrrccc}\toprule $\epsilon$ & $\ell$ & $n$ & $\omega_0$ & $\omega_{1,Z}^{\rm ADM}$ & $\omega_{1,S}^{\rm ADM}$ \\ \midrule
0.1 & 6 & 0 & $1.108459-0.087133\,\ii$ & $0.065865+0.000112\,\ii$ & $0.074345+0.000127\,\ii$ \\
0.1 & 20 & 0 & $3.596977-0.087917\,\ii$ & $0.067589+0.000003\,\ii$ & $0.076293+0.000004\,\ii$ \\
0.01 & 6 & 0 & $1.095443-0.086104\,\ii$ & $0.065051+0.000108\,\ii$ & $0.074535+0.000124\,\ii$ \\
0.01 & 20 & 0 & $3.554738-0.086885\,\ii$ & $0.066759+0.000004\,\ii$ & $0.076492+0.000004\,\ii$ \\
\bottomrule\end{tabular}\end{table}

\section{Conclusions}\label{sec:conclusions}
\begingroup
\setlength{\parskip}{0pt}
We have studied axial quasinormal modes of slowly rotating black holes surrounded by Hernquist halos, comparing two configurations of the halo matter. The dominant environmental effect on the computed low multipole modes is a common gravitational redshift. The vacuum cavity preserves the local black hole geometry, while the halo changes the relation between local time and the time measured at infinity. Consequently, appreciable shifts in both the oscillation frequency and damping rate can coexist with a spectrum whose relative mode structure remains close to that of an isolated black hole. In the dilute limit, this common shift is controlled primarily by halo compactness, explaining the weak dependence on the black-hole-to-halo mass ratio.

Rotation reveals the importance of distinguishing the angular momentum of the horizon from that of the entire spacetime. The two matter configurations can produce markedly different rotational splittings at fixed total angular momentum, yet their real spin coefficients become nearly equal when compared at fixed horizon angular momentum. Much of the apparent sensitivity to the motion of the halo therefore reflects a change in the spin carried by the central black hole. Together with the common redshift and the uniform frame dragging offset, this angular momentum partition provides a physical explanation of the leading rotational response.

The remaining spectral differences probe wave propagation through the halo beyond the local cavity description. Although small in the full complex frequencies, these corrections need not be small relative to the rotational contribution to damping. Their competition with the local Kerr response can change the compactness trend of that contribution, as seen in the Static fundamental mode. Mode ratios that cancel the common redshift and comparisons of the damping responses of different overtones thus offer complementary ways to characterize the departures from a uniformly rescaled vacuum spectrum.

These results motivate extending the comparison at fixed horizon angular momentum to finite spin and to a broader range of halo profiles. Including dynamical matter responses would further clarify how the surrounding medium modifies the relative mode structure. Such developments would help connect the leading effects of gravitational redshift and angular momentum partition with the finer spectral features that carry information about the halo.

\par
\endgroup

\begin{acknowledgments}
This work is supported by NSFC (National Natural Science Foundation of China) Grant No.~11905157 and No.~11935009, Tianjin University Self-Innovation Fund Extreme Basic Research Project Grant No.~2025XJ21-0007.
\end{acknowledgments}

\bibliography{SRDMBH}

\begin{thebibliography}{99}
\setlength{\itemsep}{0pt}
\setlength{\parsep}{0pt}
\bibitem{BertiReview2009} E. Berti, V. Cardoso, and A. O. Starinets,
``Quasinormal modes of black holes and black branes,''
\href{https://doi.org/10.1088/0264-9381/26/16/163001}{Class. Quantum Grav. \textbf{26}, 163001 (2009)}, \href{https://arxiv.org/abs/0905.2975}{arXiv:0905.2975}.
\bibitem{Spieksma2025} T. F. M. Spieksma, V. Cardoso, G. Carullo, M. Della Rocca, and F. Duque,
``Black hole spectroscopy in environments: detectability prospects,'' \href{https://doi.org/10.1103/PhysRevLett.134.081402}{Phys. Rev. Lett. \textbf{134}, 081402 (2025)}, \href{https://arxiv.org/abs/2409.05950}{arXiv:2409.05950}.
\bibitem{Leung1999} P. T. Leung, Y. T. Liu, W.-M. Suen, C. Y. Tam, and K. Young,
``Perturbative approach to the quasinormal modes of dirty black holes,''
\href{https://doi.org/10.1103/PhysRevD.59.044034}{Phys. Rev. D \textbf{59}, 044034 (1999)}, \href{https://arxiv.org/abs/gr-qc/9903032}{arXiv:gr-qc/9903032}.
\bibitem{Barausse2014} E. Barausse, V. Cardoso, and P. Pani,
``Can environmental effects spoil precision gravitational-wave astrophysics?,''
\href{https://doi.org/10.1103/PhysRevD.89.104059}{Phys. Rev. D \textbf{89}, 104059 (2014)}, \href{https://arxiv.org/abs/1404.7149}{arXiv:1404.7149}.
\bibitem{Cardoso2022} V. Cardoso, K. Destounis, F. Duque, R. P. Macedo, and A. Maselli,
``Black holes in galaxies: environmental impact on gravitational-wave generation and propagation,''
\href{https://doi.org/10.1103/PhysRevD.105.L061501}{Phys. Rev. D \textbf{105}, L061501 (2022)}, \href{https://arxiv.org/abs/2109.00005}{arXiv:2109.00005}.
\bibitem{Figueiredo2023} E. Figueiredo, A. Maselli, and V. Cardoso,
``Black holes surrounded by generic dark matter profiles: appearance and gravitational-wave emission,''
\href{https://doi.org/10.1103/PhysRevD.107.104033}{Phys. Rev. D \textbf{107}, 104033 (2023)}, \href{https://arxiv.org/abs/2303.08183}{arXiv:2303.08183}.
\bibitem{Shen2024Models} Z. Shen, A. Wang, Y. Gong, and S. Yin,
``Analytical models of supermassive black holes in galaxies surrounded by dark matter halos,''
\href{https://doi.org/10.1016/j.physletb.2024.138797}{Phys. Lett. B \textbf{855}, 138797 (2024)}, \href{https://arxiv.org/abs/2311.12259v4}{arXiv:2311.12259v4}.
\bibitem{Shen2024Inner} Z. Shen, A. Wang, and S. Yin,
``Inner radius and energy conditions of dark matter halos surrounding Schwarzschild black holes,''
\href{https://doi.org/10.1016/j.physletb.2025.139300}{Phys. Lett. B \textbf{862}, 139300 (2025)}, \href{https://arxiv.org/abs/2408.05417}{arXiv:2408.05417}.
\bibitem{Zhao2024} Y. Zhao, B. Sun, Z. Cao, K. Lin, and W.-L. Qian,
``Influence of dark matter equation of state on the axial gravitational ringing of supermassive black holes,'' \href{https://doi.org/10.1103/PhysRevD.109.044031}{Phys. Rev. D \textbf{109}, 044031 (2024)}, \href{https://arxiv.org/abs/2308.15371}{arXiv:2308.15371}.
\bibitem{Pezzella2025} L. Pezzella, K. Destounis, A. Maselli, and V. Cardoso,
``Quasinormal modes of black holes embedded in halos of matter,''
\href{https://doi.org/10.1103/PhysRevD.111.064026}{Phys. Rev. D \textbf{111}, 064026 (2025)}, \href{https://arxiv.org/abs/2412.18651}{arXiv:2412.18651}.
\bibitem{FengZhang2025} X.-H. Feng and G.-Y. Zhang,
``Shadow and quasi-normal modes of Schwarzschild--Hernquist black hole,'' \href{https://doi.org/10.1140/epjc/s10052-026-15293-z}{Eur. Phys. J. C \textbf{86}, 36 (2026)}, \href{https://arxiv.org/abs/2509.04001}{arXiv:2509.04001}.
\bibitem{QiCai2026} S. Qi and Z. Cai,
``Shadows and quasinormal modes of a Schwarzschild black hole immersed in Hernquist dark matter halo,'' \href{https://doi.org/10.1140/epjc/s10052-026-15331-w}{Eur. Phys. J. C \textbf{86}, 94 (2026)}.
\bibitem{Chakraborty2025} S. Chakraborty, G. Comp\`ere, and L. Machet,
``Tidal Love numbers and quasinormal modes of the Schwarzschild--Hernquist black hole,'' \href{https://doi.org/10.1103/4p2c-rwdh}{Phys. Rev. D \textbf{112}, 024015 (2025)}, \href{https://arxiv.org/abs/2412.14831}{arXiv:2412.14831}.
\bibitem{Chen2026} C.-Y. Chen, H. Hassanabadi, and S. Zare,
``Quasinormal-mode redshifts of black holes in galactic halos with radial pressures,'' \href{https://arxiv.org/abs/2609.27379}{arXiv:2609.27379} (2026).
\bibitem{Fernandes2025} P. G. S. Fernandes and V. Cardoso,
``Spinning black holes in astrophysical environments,'' \href{https://doi.org/10.1103/9shv-5d21}{Phys. Rev. Lett. \textbf{135}, 211403 (2025)}, \href{https://arxiv.org/abs/2507.04389}{arXiv:2507.04389}.
\bibitem{Datta2026} S. Datta and C. Singha,
``Geometric properties of slowly rotating black holes embedded in matter environments,'' \href{https://doi.org/10.1103/hln1-q5hl}{Phys. Rev. D \textbf{113}, 124052 (2026)}, \href{https://arxiv.org/abs/2602.10579}{arXiv:2602.10579}.
\bibitem{PaniSlow2012} P. Pani, V. Cardoso, L. Gualtieri, E. Berti, and A. Ishibashi,
``Perturbations of slowly rotating black holes: Massive vector fields in the Kerr metric,''
\href{https://doi.org/10.1103/PhysRevD.86.104017}{Phys. Rev. D \textbf{86}, 104017 (2012)}, \href{https://arxiv.org/abs/1209.0773}{arXiv:1209.0773}.
\bibitem{Pani2013} P. Pani, ``Advanced Methods in Black-Hole Perturbation Theory,''
\href{https://doi.org/10.1142/S0217751X13400186}{Int. J. Mod. Phys. A \textbf{28}, 1340018 (2013)}, \href{https://arxiv.org/abs/1305.6759}{arXiv:1305.6759}.
\bibitem{FengPeng2024} X.-H. Feng and J. Peng,
``Axial gravitational perturbations of slowly rotating compact objects in general relativity and beyond,''
\href{https://doi.org/10.1103/PhysRevD.110.064078}{Phys. Rev. D \textbf{110}, 064078 (2024)}, \href{https://arxiv.org/abs/2404.16437}{arXiv:2404.16437}.
\bibitem{Hernquist1990} L. Hernquist,
``An analytical model for spherical galaxies and bulges,''
\href{https://doi.org/10.1086/168845}{Astrophys. J. \textbf{356}, 359--364 (1990)}.
\bibitem{Sadeghian2013} L. Sadeghian, F. Ferrer, and C. M. Will,
``Dark-matter distributions around massive black holes: A general relativistic analysis,'' \href{https://doi.org/10.1103/PhysRevD.88.063522}{Phys. Rev. D \textbf{88}, 063522 (2013)}, \href{https://arxiv.org/abs/1305.2619}{arXiv:1305.2619}.
\bibitem{Fornberg1988} B. Fornberg,
``Generation of finite difference formulas on arbitrarily spaced grids,''
\href{https://doi.org/10.1090/S0025-5718-1988-0935077-0}{Math. Comp. \textbf{51}, 699--706 (1988)}.
\bibitem{LinQian2016} K. Lin and W.-L. Qian,
``A non grid-based interpolation scheme for the eigenvalue problem,''
\href{https://arxiv.org/abs/1609.05948}{arXiv:1609.05948} (2016).
\bibitem{LinQian2017} K. Lin and W.-L. Qian,
``A matrix method for quasinormal modes: Schwarzschild black holes in asymptotically flat and (anti-) de Sitter spacetimes,''
\href{https://doi.org/10.1088/1361-6382/aa6643}{Class. Quantum Grav. \textbf{34}, 095004 (2017)}, \href{https://arxiv.org/abs/1610.08135}{arXiv:1610.08135}.
\bibitem{Trefethen2000} L. N. Trefethen,
\href{https://doi.org/10.1137/1.9780898719598}{\textit{Spectral Methods in MATLAB} (SIAM, Philadelphia, 2000)}.
\bibitem{Jansen2017} A. Jansen,
``Overdamped modes in Schwarzschild-de Sitter and a Mathematica package for the numerical computation of quasinormal modes,''
\href{https://doi.org/10.1140/epjp/i2017-11825-9}{Eur. Phys. J. Plus \textbf{132}, 546 (2017)}, \href{https://arxiv.org/abs/1709.09178}{arXiv:1709.09178}.
\bibitem{Cardoso2009} V. Cardoso, A. S. Miranda, E. Berti, H. Witek, and V. T. Zanchin,
``Geodesic stability, Lyapunov exponents, and quasinormal modes,''
\href{https://doi.org/10.1103/PhysRevD.79.064016}{Phys. Rev. D \textbf{79}, 064016 (2009)}, \href{https://arxiv.org/abs/0812.1806}{arXiv:0812.1806}.
\bibitem{HatsudaKimura2020} Y. Hatsuda and M. Kimura,
``Semi-analytic expressions for quasinormal modes of slowly rotating Kerr black holes,''
\href{https://doi.org/10.1103/PhysRevD.102.044032}{Phys. Rev. D \textbf{102}, 044032 (2020)}, \href{https://arxiv.org/abs/2006.15496}{arXiv:2006.15496}.
\end{thebibliography}
\end{document}